\documentclass[12pt, preprint]{aastex}
\shorttitle{The Dark Side of ROTSE-III Prompt GRB Observations}
\shortauthors{Yost {\it et al.}}

\def\lsim{\mathrel{\rlap{\lower4pt\hbox{\hskip1pt$\sim$}}
    \raise1pt\hbox{$<$}}}                
\def\gsim{\mathrel{\rlap{\lower4pt\hbox{\hskip1pt$\sim$}}
    \raise1pt\hbox{$>$}}}                
\begin{document}

\title{The Dark Side of ROTSE-III Prompt GRB Observations}

\author{
Yost,~S.~A.\altaffilmark{1},
Aharonian,~F.\altaffilmark{2},
Akerlof,~C.~W.\altaffilmark{1},
Ashley,~M.~C.~B.\altaffilmark{3}, 
Barthelmy, S.\altaffilmark{4},
Gehrels, N.\altaffilmark{4},
G\"o\u{g}\"u\c{s},~E.\altaffilmark{5},
G\"{u}ver, T.\altaffilmark{6},
Horns,~D.\altaffilmark{2},
K{\i}z{\i}lo\v{g}lu,~\"{U}.\altaffilmark{7},
Krimm, H. A.\altaffilmark{4,8}
McKay,~T.~A.\altaffilmark{1},
\"{O}zel,~M.\altaffilmark{9},
Phillips,~A.\altaffilmark{3}, 
Quimby,~R.~M.\altaffilmark{10},
Rowell, G.\altaffilmark{2,11},
Rujopakarn,~W.\altaffilmark{12},
Rykoff,~E.~S.\altaffilmark{1}, 
Schaefer,~B.~E.\altaffilmark{13}, 
Smith,~D.~A.\altaffilmark{1,14},
Swan,~H.~F.\altaffilmark{1},
Vestrand,~W.~T.\altaffilmark{15},
Wheeler,~J.~C.\altaffilmark{10},
Wren,~J.\altaffilmark{15},
Yuan,~F.\altaffilmark{1}
}

\altaffiltext{1}{University of Michigan, 2477 Randall Laboratory, 450
        Church St., Ann Arbor, MI, 48104, sayost@umich.edu, 
         akerlof@umich.edu,
        tamckay@umich.edu, erykoff@umich.edu, donaldas@umich.edu,hswan@umich.edu,
 yuanfang@umich.edu}
\altaffiltext{2}{Max-Planck-Institut f\"{u}r Kernphysik, Saupfercheckweg 1,
        69117 Heidelberg, Germany, Felix.Aharonian@mpi-hd.mpg.de,
        horns@mpi-hd.mpg.de, rowell@mpi-hd.mpg.de}
\altaffiltext{3}{School of Physics, Department of Astrophysics and Optics,
        University of New South Wales, Sydney, NSW 2052, Australia,
        mcba@phys.unsw.edu.au, a.phillips@unsw.edu.au}
\altaffiltext{4}{NASA/Goddard Space Flight Center, Greenbelt, MD 20771, scott@lheamail.gsfc.nasa.gov, gehrels@gsfc.nasa.gov, krimm@milkyway.gsfc.nasa.gov}
\altaffiltext{5}{Sabanc{\i} University, Orhanl{\i}-Tuzla 34956 Istanbul, Turkey, ersing@sabanciuniv.edu}
\altaffiltext{6}{Istanbul University Science Faculty, Department of Astronomy
        and Space Sciences, 34119, University-Istanbul, Turkey, 
        tolga@istanbul.edu.tr}
\altaffiltext{7}{Middle East Technical University, 06531 Ankara, Turkey,
        umk@astroa.physics.metu.edu.tr}
\altaffiltext{8}{Universities Space Research Association, 10211 Wincopin Circle, Suite 500, Columbia, MD  21044-3432, krimm@milkyway.gsfc.nasa.gov}
\altaffiltext{9}{\c{C}anakkale Onsekiz Mart \"{U}niversitesi, Terzio\v{g}lu
        17020, \c{C}anakkale, Turkey, m.e.ozel@comu.edu.tr}
\altaffiltext{10}{Department of Astronomy, University of Texas, Austin, TX
        78712, quimby@astro.as.utexas.edu, wheel@astro.as.utexas.edu}
\altaffiltext{11}{School of Chemistry \& Physics, University of Adelaide, Adelaide 5005, Australia,rowell@mpi-hd.mpg.de}
\altaffiltext{12}{Steward Observatory
Tucson, AZ, 85721, wiphu@as.arizona.edu }
\altaffiltext{13}{Department of Physics and Astronomy, Louisiana State
        University, Baton Rouge, LA 70803, schaefer@lsu.edu}
\altaffiltext{14}{Guilford College, 5800 West Friendly Ave., Greensboro, NC 27410, dsmith4@guilford.edu}
\altaffiltext{15}{Los Alamos National Laboratory, NIS-2 MS D436, Los Alamos, NM 87545, vestrand@lanl.gov, jwren@nis.lanl.gov}

\begin{abstract}

We present several cases of optical observations during $\gamma$-ray
bursts (GRBs) which resulted in prompt limits but no detection of
optical emission. These limits constrain the prompt optical flux
densities and the optical brightness relative to the $\gamma$-ray
emission. The derived constraints fall within the range of properties
observed in GRBs with prompt optical detections, though at the faint
end of optical/$\gamma$ flux ratios. The presently accessible prompt
optical limits do not require a different set of intrinsic or
environmental GRB properties, relative to the events with prompt
optical detections.

\end{abstract}

\section{Introduction}

Since the launch of the {\it Swift} satellite \citep{gcgmn04}, early
long-wavelength observations of $\gamma$-ray bursts (GRBs) have become
routine.  {\it Swift} has provided prompt triggers to events since
early 2005, for which ``prompt'' signifies ``during $\gamma$-ray
emission''. There is a growing number of optical lightcurves that
begin during, or within seconds after, the $\gamma$-ray
emission. There are also several cases with prompt optical
non-detections which constrain the optical brightness during the GRB.

Prompt and very early broadband emission has been the major advance in
{\it Swift}-era GRB studies, opening serious investigations of
important physical questions. One example is the nature of the
relativistic outflow, generally thought of as baryonic with energy
released by internal shocks. The proposed alternatives include
magnetized flows which release energy via magnetic reconnection
\citep{mrp94,thompson94,usov94}. The early broadband detections at
X-ray and optical wavelengths are now being used to test these models
\citep[e.g.,][]{kmpwo07}.

From the beginning of the afterglow discovery era, optical
counterparts have been found to have a large range in
brightness. Despite good observations, a significant fraction
($\sim$\,50\%) of events do not have detected optical
afterglows. These ``optically dark'' GRBs have produced questions
regarding GRB physics and environment \citep[see pre- and post-{\it
Swift} reviews, such as][respectively]{p05,Zhang07}. 

Nondetections during prompt optical observations are not precisely the
same as these optically dark GRBs. In a few events deeper post-GRB
observations detect the optical transient. This raises the question as
to whether prompt limits are ``promptly dark''; are the limiting
fluxes consistent with the brightness range observed in prompt optical
detections, or do prompt nondetections require a separate population
of optical properties? Such properties could be due to either
intrinsic (faint events, or faint optical--to--$\gamma$-ray flux
ratios) or extrinsic (local dust absorption, or the Lyman-$\alpha$
forest absorption from high $z$) causes. 

``Excessively'' faint prompt optical emission would therefore have
interesting implications for the GRB spectral shape or
environment. While the peak frequency of the GRB has often been
constrained \citep[$\nu$$f_{\nu}$ peaking near a few 100 keV, see the
review by][]{p05}, the shape of the prompt emission's low-energy tail
is not well known, with self-absorption frequency estimates from the
optical to X-ray \citep[e.g.,][]{pw04,wei07}. As well, indications of
high redshift would be important. While there are suggested redshift
indicators from GRB $\gamma$-ray properties alone, these are not
proven, as discussed critically by \citet{bkbc07}.

The ROTSE-III project has provided some of the earliest optical
observations of GRB triggers, with a number of detections. To date, there
has been no consistent correlation between prompt optical fluxes
and the contemporaneous $\gamma$-rays \citep[e.g., see the discussions
in][]{rykaa05,ysraa07}. This paper discusses prompt ROTSE-III
observations under good sky conditions which did not yield
detections. The limits placed upon the ratio of optical emission to
the higher energy emission are discussed in comparison with the
behavior associated with prompt detections.

In the following discussion, the spectral flux density is
characterized by the spectral index $\beta$, with $f_{\nu} \propto
\nu^{\beta}$.  This convention relates $\beta$ to the $\gamma$-ray
photon index $\Gamma$ by $\beta = 1 - \Gamma$.  To designate a
spectral region, subscripts ``OPT'', ``X'', and ``$\gamma$'' for
$\beta$ indicate an index for the optical, X-ray, and $\gamma$-ray
bands respectively. A spectral index spanning two regions is
designated with both, e.g., $\beta_{\mathrm{OPT}-\gamma}$ for the
spectral index interpolating between the optical and $\gamma$-ray
frequencies. 

We note briefly that the overall spectral and temporal shape of
afterglows typically suggests synchrotron emission from a fireball whose
accelerated electrons have a Lorentz factor distribution
$N(\gamma_e) \propto \gamma_e^{-p}$ \citep[this is reviewed, e.g.,
by][]{m06}. The afterglow spectrum has spectral breaks: principally
$\nu_m$, due to the minimum Lorentz factor $\gamma_e$, and $\nu_c$,
the cooling frequency. These provide predictions for the spectral
shape of a single synchrotron component. The index $\beta = 1/3$ at
frequencies below the peak in $f_{\nu}$ ($\nu < \nu_m$), $\beta =
(1-p)/2$ for $\nu_m < \nu < \nu_c$, and $\beta = -p/2$ for the case
when $\nu > \nu_c$ and $\nu > \nu_m$. (When $\nu_c < \nu_m$, the
spectral shape is $\nu^{-1/2}$ for frequencies between them.) These
predictions, with $\beta$ from $1/3$ to $-3/2$ for $p = 2$ -- $3$,
 can be compared to the constraints upon
$\beta_{\mathrm{OPT}-\gamma}$.

\begin{figure}
\epsscale{0.8}
\plotone{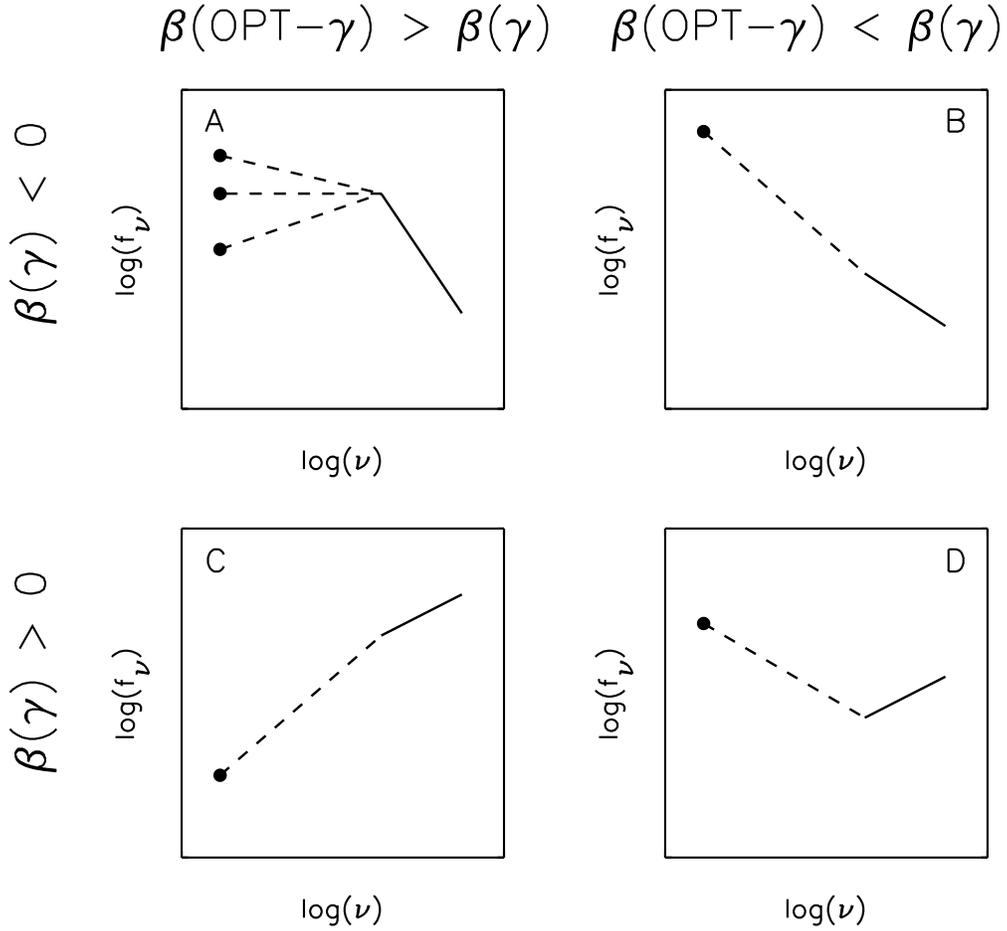}
\caption{Diagrams representing some possible optical--to--$\gamma$-ray
spectra. These illustrate information available from the spectral
indices $\beta_{\gamma}$ (solid lines) and
$\beta_{\mathrm{OPT}-\gamma}$ (dashed lines). In nearly all observed
GRBs, the spectrum at the lower-energy $\gamma$-rays (BAT band) has
$\beta_{\gamma} < 0$, as in the two upper panels. Rarely,
$\beta_{\gamma} > 0$ (lower panels). The comparison of the two $\beta$
constrains whether the $\gamma$-ray spectrum over- (left, with
$\beta_{\mathrm{OPT}-\gamma} > \beta_{\gamma}$) or under- (right, with
$\beta_{\mathrm{OPT}-\gamma} < \beta_{\gamma}$) predicts the optical
flux. With optical limits, an underprediction by $\beta_{\gamma}$
(optical excess) cannot be inferred, but an overprediction can be
deduced; spectra allowing (B) or (A) can be differentiated from cases
which only allow (A), and those congruent with (D) or (C) from those
which only permit (C).}\label{fig:schema}
\end{figure}

Figure \ref{fig:schema} shows some possible combinations of
$\beta_{\mathrm{OPT}-\gamma}$ and $\beta_{\gamma}$. The $\gamma$-ray
spectrum may predict the optical flux ($\beta_{\mathrm{OPT}-\gamma} =
\beta_{\gamma}$), indicating that a single power-law
(synchrotron-like) component could account for the broadband
spectrum. When $\beta_{\mathrm{OPT}-\gamma} < \beta_{\gamma}$, the
$\gamma$-ray spectrum underpredicts the optical flux, implying a
separate low-energy emission component. When
$\beta_{\mathrm{OPT}-\gamma} > \beta_{\gamma}$, the $\gamma$-ray
spectrum overpredicts the optical flux, indicating a spectral rollover
between the optical and high frequencies. When there are only prompt
optical upper limits in flux density, one can nevertheless
discriminate between cases where $\beta_{\gamma}$ either predicts or
overpredicts the optical flux limit from those where $\beta_{\gamma}$
could underpredict the optical flux.

\section{Optical Observations}

The ROTSE-III array is a worldwide network of 0.45~m robotic, automated
telescopes, built for fast ($\sim 6$ s) responses to GRB triggers from
satellites such as {\it Swift}.  They have a wide ($1\fdg85 \times 1\fdg85$) field
of view imaged onto a Marconi $2048\times2048$ back-illuminated thinned CCD,
and operate without filters. The ROTSE-III systems are described in detail in
\citet{akmrs03}.

ROTSE-III images were reduced and processed using the RPHOT pipeline,
with routines based upon DAOPHOT \citep{stetson87}. Objects were
identified via SExtractor \citep{ba96} and calibrated astrometrically
and photometrically with the USNOB1.0 catalog. They are tied to the
$R$ band, and these unfiltered ``$R$-equivalent'' magnitudes are
designated as ``$C_R$''. The method is fully described in
\citet{qryaa05}. The final result yields limiting magnitudes in the
GRB error box from the PSF-fit photometric data. These are presented
in Table \ref{tab:optlim}.

\subsection{Sources of Prompt Detection Data}\label{sec:detdata}

Table \ref{tab:optgam1} presents spectral index information for
several GRBs with prompt optical detections. These are used to provide
a comparison for prompt limit results. The table is similar to Table 5
of \citet{ysraa07}, which is also used for comparison. 

Most of the prompt optical detections used for this table are from
ROTSE-III observations. These include GRB\,060111B \citep{GCN4488},
GRB\,060729 \citep{GCN5366}, GRB\,060904B \citep{GCN5504} and
GRB\,061007 \citep{GCN5706} which are discussed in a comprehensive
analysis paper (Rykoff {\it et al.}, in prep.). GRB\,061121 was
promptly detected by ROTSE; these data are presented in
\citet{page07}. GRB\,060927 was a high-redshift event
\citep{GCN5651}. The prompt ROTSE detection is converted to a flux
density at a wavelength near $i$-band, as described in Ruiz-Velasco
{\it et al.}, in prep. GRB\,060218 was detected by ROTSE (Table
\ref{tab:optgam1}: line with a $C_R$ observation of GRB\,060218)
\citep{GCN4782} and by the {\it Swift} UVOT (Table \ref{tab:optgam1}:
line with a $V$ observation of GRB\,060218) \citep{cmbbb06}. Finally,
GRB\,050820A and GRB\,061126 were promptly detected by RAPTOR. For the
former, we determine optical and $\gamma$-ray flux densities from
\citet{vwwag06}. For the latter, we take the optical flux densities of
\citet{pbbph07}, correcting for Galactic extinction.

\subsection{Prompt Nondetections with Later Detections}\label{latedet}

One case of a prompt limit with a later detection is the first ROTSE
observation of GRB\,060729. The OT flux was rising, and the second
5-second image was the first to yield a detection (Tables
\ref{tab:optgam1} and \ref{tab:optgam2} show that the flux rises from
$\sim$ 1/2 to 2 mJy over the first few images). 

In GRB\,060614, the ROTSE limits at 29 seconds post-trigger were
obtained before the subsequent UVOT afterglow detection at 100 seconds
post-trigger. This initial UVOT V-band detection \citep{GCN5252} had
notable flux uncertainty (18.4$\pm$0.5 mag) but is significantly
(nearly 3 mag) fainter than the ROTSE limits. The ROTSE limit values
are fully consistent with the later detection, and constrain the flux
decay to have been no more rapid than $\sim t^{-2}$ from a half minute
to two minutes post-trigger. 

Optical detections indicate that the GRB cannot be at high $z$, as the
Lyman-$\alpha$ forest would absorb the optical flux. Indeed,
GRB\,060729 has $z = 0.54$ \citep{GCN5373}, and the host of
GRB\,060614 is at $z = 0.125$ \citep{GCN5275}, \citep[although there
is some controversy, with an estimate of $z \approx 1.5$,][]{sx06}.
There are further GRBs with prompt limits followed by optical
detections at $t \gsim$\,1\,hr: GRB\,050306 \citep{GCN3089},
GRB\,050713A \citep{GCN3582}, and GRB\,061110 \citep{GCN5797}. The
prompt nondetections of these three events cannot be attributed to
high $z$.

\section{High Energy Data}\label{sec:hidata}

BAT data were used for $\gamma$-ray comparisons in these {\it Swift}
bursts. For the $\gamma$-ray data, the event files from the public
archives were analyzed with the BATTOOLS and XSPEC11 software
packages\footnote{{\tt
http://swift.gsfc.nasa.gov/docs/swift/analysis/}} . The result is
unabsorbed flux values in the 15--150~keV range. When there is
sufficient signal (for $\approx$\,30\% of the data points), these are
determined directly along with $\beta_{\gamma}$ during the precise
time interval of each optical observation. For the remainder of the
data where the signal is insufficient, the count rate during the
interval is converted to fluxes using the BAT spectrum during a
longer, overlapping interval. The analyses are the same as described
for the GRB\,051109A and GRB\,051111 events, in \citet{ysraa07}.

In addition, a few events have prompt X-ray data (in the $1-10$ keV
band, with an effective frequency $\nu < 10^{18}$\,Hz) as well. Table
\ref{tab:optgam1} lists results with simultaneous optical, X-ray, and
$\gamma$-ray detections for GRB\,060729, GRB\,060904B, and
GRB\,061007. The XRT analyses are fully discussed in an upcoming ROTSE
paper treating multiband lightcurves (Rykoff {\it et al.}, in
prep.). In brief, the {\tt xrtpipeline} tool calibrates and performs
standard filtering and screening. This is followed by count
extractions from appropriate regions for the source and background,
the generation of response files with the FTOOLS task {\tt xrtmkarf},
and spectral fits to yield fluxes. For GRB\,061121, the XRT data is
taken directly from the flux densities in \citet{page07}; the
reductions were similar and compensate for the significant pileup
effects, as discussed there in detail.

There are two cases with optical nondetections and X-ray prompt
detections within the sample presented, GRB\,050713A and
GRB\,060614. A limit upon $\beta_{\mathrm{OPT}-X}$ adds little to the
information from the $\beta_{\mathrm{OPT}-\gamma}$ limit; the events
are compatible with an interpretation of the prompt $t \approx
100$~sec X-ray flux as an extension of the contemporaneous
$\gamma$-rays.  This was seen in a quick analysis of the GRB\,060614
archive data \citep[as well as the spectral information given in
][]{GCN5254, GCN5256}, and by the \citet{owogpv06} analysis of
GRB\,050713A XRT and BAT data. Further detailed comparisons are beyond
the scope of this paper.

\section{Determining $\beta_{\mathrm{OPT}-\gamma}$ and $\beta_{\mathrm{OPT}-X}$}

The spectral index (or its limit) was determined between the optical
and higher-energy bands in the same manner as those presented in
\citet{ysraa07}. In brief, the optical data was corrected for Galactic extinction
and converted to flux densities as if the $C_R$ magnitudes were $R$,
using the zeropoints of \citet{b79}. These data are in Tables
\ref{tab:optgam1}, \ref{tab:optgam2}, along with the flux densities of
the $\gamma$-ray detections (and X-ray, where applicable). The flux
densities and effective frequencies of the bands are then used to
calculate $\beta$. When the optical is not detected, the optical limit
is used with the lower (1\,$\sigma$) estimate of the high-energy
emission to estimate the softest spectral index
$\beta_{\mathrm{OPT}-\gamma}$ (or $\beta_{\mathrm{OPT}-X}$) possible.

The Galactic extinction corrections are taken from
\citet{sfd98}. $C_R$ limits are treated as $R$-equivalent and adjusted
for the $R$ band's extinction.  Table \ref{tab:optgam1} gives the flux
and $\beta$ results for cases with prompt optical detections, in the
same manner as \citet{ysraa07}. Table \ref{tab:optgam2} gives flux and
$\beta$ constraints for events with prompt optical limits.

\section{Discussion}

We consider 27 GRBs with prompt optical observations, the data
presented in Tables \ref{tab:optgam1} and \ref{tab:optgam2}, as well
as Table 5 in \citet{ysraa07}. 11 of these GRBs only had prompt
optical limits, while 14 were consistently promptly detected in the
optical, and a further 2 events had both prompt limits and
detections. The data includes a total of 43 distinct prompt optical
detections, and 55 prompt optical limits.

\subsection{Diverse Prompt Properties}

\begin{figure}
\epsscale{1.0}
\plotone{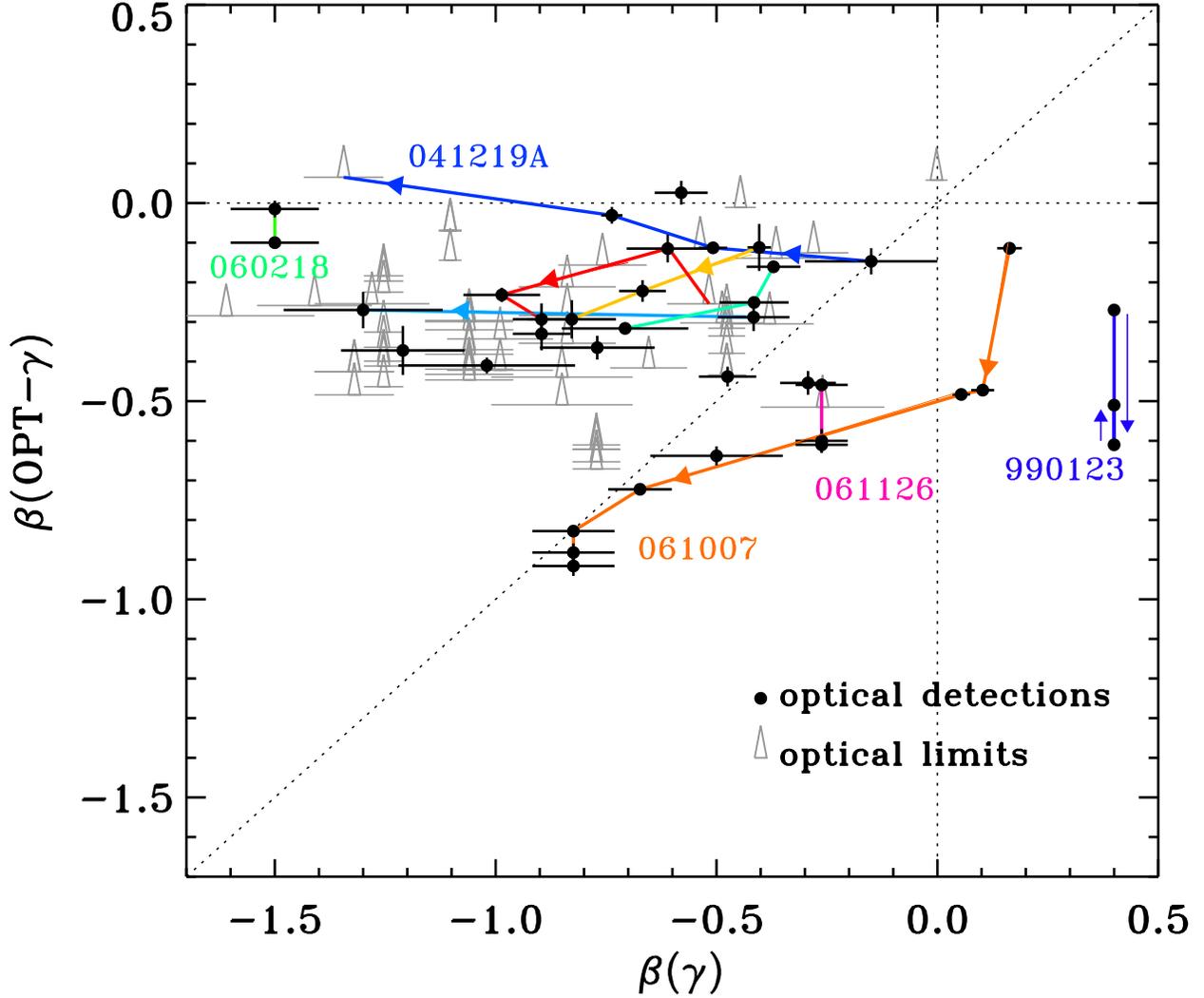}

\caption{
Optical-to-$\gamma$ spectral indices ($\beta_{\mathrm{OPT}-\gamma}$)
plotted against $\gamma$-ray spectral indices ($\beta_{\gamma}$). Data
are from Tables \ref{tab:optgam1} and \ref{tab:optgam2}, as well as
Table 5 in \citet{ysraa07}. Optical detections have black points, and
cases with optical limits are grey triangles. The latter indicate the
{\it softest} possible $\beta_{\mathrm{OPT}-\gamma}$; the triangles
are used to point upwards instead of arrows. When several optical
detections occur in a single GRB event, the points are connected by a
(colored in online version) line. Where legible, an arrow points from
earlier to later observations.\newline
\newline
Most cases are above the $\beta_{\mathrm{OPT}-\gamma} =
\beta_{\gamma}$ line, with $\beta_{\mathrm{OPT}-\gamma} >
\beta_{\gamma}$. The $\gamma$-ray spectrum overpredicts the optical
flux; this indicates a spectral rollover between the optical and high
frequencies, whether or not there are separate emission components at
optical and $\gamma$-ray energies. Sometimes
$\beta_{\mathrm{OPT}-\gamma} < \beta_{\gamma}$. The $\gamma$-ray
spectrum underpredicts the optical flux, implying a separate
low-energy emission component. A few cases have consistent indices,
which, as discussed for GRB\,051111 \citep{ysraa07}, could indicate a
single spectral shape extending from $\gamma$-ray to optical
energies. The optical limits are consistent with
$f_{\nu}$(OPT)/$f_{\nu}$($\gamma$) ratios from optical detections and
do not imply a separate population whose prompt optical emission is
fainter relative to the $\gamma$-rays.
}\label{fig:betas}

\end{figure}

GRBs show diversity in their prompt optical and $\gamma$-ray
brightnesses. Optical flux densities span 100 $\mu$Jy to 3 Jy while
contemporaneous $\gamma$-ray flux densities take values from 6 $\mu$Jy
to 4 mJy. This results in a range of possible prompt spectral indices
$\beta_{\mathrm{OPT}-\gamma}$ and $\beta_{\gamma}$, which are plotted
as $\beta_{\mathrm{OPT}-\gamma}$ against $\beta_{\gamma}$ in Figure
\ref{fig:betas}.

In this dataset, all relations between $\beta_{\mathrm{OPT}-\gamma}$
and $\beta_{\gamma}$ are observed ($\beta_{\mathrm{OPT}-\gamma} >
\beta_{\gamma}$, $\beta_{\mathrm{OPT}-\gamma} < \beta_{\gamma}$, or
$\beta_{\mathrm{OPT}-\gamma} = \beta_{\gamma}$).  The values of
$\beta_{\mathrm{OPT}-\gamma}$ and $\beta_{\gamma}$ vary widely, from
$-0.9$ to $0.03$ for $\beta_{\mathrm{OPT}-\gamma}$ and from $-1.5$ to
0.4 for $\beta_{\gamma}$. These are within or quite close to the range
of $\beta = 1/3$ to $-3/2$ for the synchrotron spectral shape
(discussed in the Introduction) and electron energy distribution
indices of $p = 2$ -- $3$.

In addition, observations of some events show both
$\beta_{\mathrm{OPT}-\gamma}$ and $\beta_{\gamma}$ changing
significantly during a burst. $\beta_{\gamma}$ generally evolves from
hard to soft. This is a previously known characteristic of many GRBs
\citep[e.g., as reviewed by][]{fm95}, now considered in models of
prompt emission \citep[such as ``jitter'' radiation;][]{medved06}.  The
changes in $\beta_{\mathrm{OPT}-\gamma}$ indicate that optical prompt
fluxes are generally not correlated with the $\gamma$-ray emission.

There has been discussion in the literature concerning whether prompt
optical emission is an extension of the $\gamma$-rays, or is a
separate component. \citet{vwwfs05} indicates an optical component
correlated to the GRB in GRB\,041219A, while \citet{vwwag06} and
\citet{ysraa07} discuss the apparent blend of $\gamma$-ray--correlated
and uncorrelated components in the prompt optical lightcurves of GRBs
050820A and 051111 respectively. The correlated component of
GRB\,051111 is one of the few cases where the indices allow
$\beta_{\mathrm{OPT}-\gamma} = \beta_{\gamma}$. Several events had
prompt optical behavior distinct from that of the GRB, and apparently
connected to the afterglow; the prompt optical lightcurves of
GRB\,050401, GRB\,051109A and GRB\,061126 are decaying \citep{rykaa05,
ysraa07, pbbph07}, and that of GRB\,060729 is rising
\citep{GCN5377}. There are also events where the optical flux does not
rise until after the GRB \citep[e.g., GRB\,030418, GRB\,060605, GRB
060607A, ][ respectively]{rspaa04,srsq06,GCN5236}.

As seen by the variety of $\beta_{\mathrm{OPT}-\gamma}$, there is no
universal ratio $f_{\nu}$(OPT)/$f_{\nu}$($\gamma$). There is no common
$\beta_{\mathrm{OPT}-\gamma}$ / $\beta_{\gamma}$ connection in all
events, but in most cases, $\beta_{\mathrm{OPT}-\gamma}$ is harder
than $\beta_{\gamma}$. For these, $\beta_{\gamma}$ overpredicts the
optical, requiring a rollover in the spectrum between the $\gamma$-ray and
optical frequencies, whether or not there are separate emission components
at optical and $\gamma$-ray energies. Nearly all the limits give
$\beta_{\mathrm{OPT}-\gamma}$ versus $\beta_{\gamma}$ falling into
this category.

In some prompt detections, $\beta_{\mathrm{OPT}-\gamma} <
\beta_{\gamma}$ and  $\beta_{\gamma}$ underpredicts the optical
\citep[e.g., see][for GRB\,051111 and GRB061126 respectively]{ysraa07,
pbbph07}.  This implies a separate low-energy emission component. All
the prompt limits presented exclude this possibility, except for
GRB\,060515. Its constraints are insufficient and allow either
$\beta_{\mathrm{OPT}-\gamma} < \beta_{\gamma}$ or
$\beta_{\mathrm{OPT}-\gamma} > \beta_{\gamma}$.

\subsection{Properties of Limits vs Detections}

The optical limits are not demonstrably the result of abnormally faint
prompt optical flux. The prompt flux limits are typically 16th or 17th
magnitude ($< 1$~mJy). Prompt detections have been recorded from small
fractions of a mJy to a few Jy.  As well, the $\gamma$-ray flux
densities of GRBs with prompt optical limits are similar to the lower
values of $f_{\nu}$($\gamma$) from GRBs with optical detections; both
sets of events have $f_{\nu}$($\gamma$) ranging from several $\mu$Jy
to over a mJy. The prompt limits require neither intrinsically fainter
emission nor excess absorption from dust or (high-$z$)
Lyman-$\alpha$. High redshifts are not a general solution for the
prompt optical limits, as some events are detected later (\S\ref{latedet}).

The values of GRB $\beta_{\gamma}$ contemporaneous with prompt optical
limits are similar to the $\beta_{\gamma}$ when prompt observations
gave optical detections. The $\beta_{\gamma}$ of optical nondetections
are on average softer than the $\beta_{\gamma}$ of detections, ranging
from $-1.6$ to 0, as compared to $-1.5$ to $0.4$. However, the data
are not consistently sampled, leading to no strong conclusions other
than that prompt observations of the few GRBs with the hardest
$\beta_{\gamma}$ have yielded detections rather than limits.

Similarly, the limits on $\beta_{\mathrm{OPT}-\gamma}$ for
nondetections ($> -0.7$) are in the range of most of the
$\beta_{\mathrm{OPT}-\gamma}$ from the prompt detections (from $-0.9$
to $0.03$). The prompt nondetections are consistent with coming from
the harder end of the $\beta_{\mathrm{OPT}-\gamma}$ distribution, as
all of the $\beta_{\mathrm{OPT}-\gamma}$ limits are harder than the
softest $\beta_{\mathrm{OPT}-\gamma}$ value calculated from prompt
detections. However, there is no evidence of bimodality of
$\beta_{\mathrm{OPT}-\gamma}$. It is only in one event (GRB\,061007)
that the $\beta_{\mathrm{OPT}-\gamma}$ of optical detections is softer
than the softest allowed $\beta_{\mathrm{OPT}-\gamma}$ from a prompt
limit. This was for epochs at the end of the event, which may be the
beginning of the afterglow, as $\gamma$-ray, X-ray and optical
frequencies lie on a single spectral powerlaw. There is not one set of
$\beta_{\mathrm{OPT}-\gamma}$ for the optically detected and another
for the nondetected cases. These overlaps in
$\beta_{\mathrm{OPT}-\gamma}$ and $\beta_{\gamma}$ are readily seen in
Figure \ref{fig:betas}.

\subsection{Cases with Prompt X-ray Data}

For GRB\,061121, the comparisons of $\beta_{\mathrm{OPT}-\gamma}$ and
$\beta_{\mathrm{OPT}- X}$ require a peak in the broadband $f_{\nu}$
spectrum. This is discussed in detail by \citet{page07}, where it can
be inferred to be near 1 keV initially and to subsequently drop in
frequency. GRB\,060729 also implies a peak between the optical and
X-ray during the epoch with optical, X-ray and $\gamma$-ray data. In
that case, $\beta_{\gamma}$ appears to be harder than $\beta_{X}$, but
this may be due to the general softening trend of $\beta_{\gamma}$ and
the measurement of $\beta_{\gamma}$ over the X-ray epoch using data
beginning well before the X-ray observations. A ``convex'' overall
X-ray--$\gamma$-ray spectral shape cannot be inferred from the weak
$\gamma$-ray detection.

In contrast, the GRB\,061007 prompt X-ray epochs do not demonstrate
such a peak. From the first GRB\,061007 epoch with X-ray data, the
spectral indices show that the $\gamma$-ray and X-ray bands are in a
single spectral segment. This is not unusual; \S\ref{sec:hidata}
indicates that in the two prompt optical limit cases, the X-ray and
$\gamma$-ray data could be from the same spectral segment. In
GRB\,061007, allowing for local extinction corrections, the entire
broadband spectrum (optical, X-ray, $\gamma$-ray) forms a single
spectral segment \citep[see ][Figure 2, which fits an absorbed
$\nu^{-1}$ spectrum]{mmgks07}. This would be expected for an early
afterglow where the high-energy emission from the forward shock
extends above the X-rays.

\section{Conclusion}

Prompt optical limits fall within the range of optical fluxes and
optical--to--$\gamma$-ray flux ratios observed from prompt optical
detections. The prompt limits yield constraints upon
optical--to--$\gamma$-ray flux ratios at the faint end of the ratios
measured from prompt detections.  This does not imply a different set
of intrinsic or environmental properties for events with detections
and nondetections; there is wide overlap in fluxes and flux ratios
between the limits and detections. Moreover, prompt detections show
great variety, and demonstrate diverse connections (or lack thereof)
with the contemporaneous $\gamma$-rays.

The most economical explanation for prompt optical nondetections is
that they are events drawn from the faint end of the range of prompt
optical emission. These faint counterparts are not always accessible
with the sensitivities of the small telescopes providing the bulk of
prompt responses.

\acknowledgements{ This work has been supported by NASA grants
NNG-04WC41G and F006794, NSF grants AST-0119685, 0105221 and 0407061,
the Australian Research Council, the University of New South Wales,
and the University of Michigan. JCW is supported by NSF Grant
AST-0406740.  Work performed at LANL is supported through internal
LDRD funding.}

\newcommand{\noopsort}[1]{} \newcommand{\printfirst}[2]{#1}
  \newcommand{\singleletter}[1]{#1} \newcommand{\switchargs}[2]{#2#1}

\begin{deluxetable}{lccc}
\tablewidth{0pt}
\tablecaption{Prompt Optical Limits\label{tab:optlim}}
\tabletypesize{\scriptsize}
\tablehead{
  \colhead{GRB} &
  \colhead{$t_{\mathrm{start}}$ (s)} &
  \colhead{$t_{\mathrm{end}}$ (s)} &
  \colhead{Magnitude}
}
\startdata
050306 & 64.8 & 69.8 &  $>$ 15.5 \\
050306 & 78.9 & 83.9 &  $>$ 15.8 \\
050306 & 93.5 & 98.5 &  $>$ 15.8 \\
050306 & 108.3 & 113.3 &  $>$ 15.7 \\
050306 & 122.7 & 127.7 &  $>$ 15.7 \\
050306 & 137.2 & 142.2 &  $>$ 15.8 \\
050306 & 151.8 & 156.8 &  $>$ 15.8 \\
050306 & 166.1 & 171.1 &  $>$ 15.7 \\
050306 & 180.4 & 185.4 &  $>$ 15.8 \\
\hline
050713A & 72.1 & 77.1 & $> 16.5$ \\
050713A & 104.7 & 124.7 & $> 17.2$ \\
\hline
050822 & 31.8 & 36.8 & $>$15.6  \\
050822 & 39.8 & 44.8 & $>$15.5  \\
050822 & 47.8 & 52.8 & $>$15.5 \\
050822 & 55.9 & 60.9 & $>$15.5 \\
050822 & 63.9 & 68.9 & $>$15.5 \\
050822 & 95.9 & 100.9 & $>$15.6 \\
\hline
050915A & 42.9 & 47.9 & $> 17.0$ \\
\hline
050922B & 258.6 & 263.6 & $> 16.4$ \\
050922B & 273.3 & 278.3 & $> 16.5$ \\
\hline
051001 & 85.7 & 90.7 & $> 16.3$ \\
051001 & 100.1 & 105.1 & $> 16.3$ \\
051001 & 114.3 & 119.3 & $> 16.2$ \\
051001 & 128.6 & 133.6 & $> 16.3$ \\
051001 & 143.1 & 148.1 & $> 16.3$ \\
051001 & 157.6 & 162.6 & $> 16.2$ \\
051001 & 172.3 & 177.3 & $> 16.1$ \\
051001 & 186.9 & 191.9 & $> 16.2$ \\
\hline
060312 & 20.3 & 25.3 & $> 14.1$ \\
060312 & 27.4 & 32.4 & $> 14.1$ \\
060312 & 34.4 & 39.4 & $> 14.2$ \\
060312 & 41.5 & 46.5 & $> 14.3$ \\
060312 & 48.7 & 53.7 & $> 14.3$ \\
\hline
060515 & 58.5 &  63.8 & $> 14.5$ \\
\hline
060614 & 26.8 & 31.8  & $> 15.7$ \\
060614 & 40.6 & 45.6  & $> 15.6$ \\
060614 & 55.2 & 60.2  & $> 15.6$ \\
060614 & 69.6 & 74.6  & $> 15.6$ \\
060614 & 83.9 & 88.9  & $> 15.6$ \\
060614 & 98.3 & 103.3  & $> 15.6$ \\
060614\tablenotemark{a} & 112.6 & 117.6 & $> 15.6$ \\
060614\tablenotemark{a} & 126.8 & 131.8 & $> 15.6$ \\
060614\tablenotemark{a} & 140.7 & 145.7 & $> 15.6$ \\
060614\tablenotemark{a} & 155.2 & 160.2 & $> 15.6$ \\
060614\tablenotemark{a} & 169.2 & 189.2 & $> 16.2$ \\
\hline
060729\tablenotemark{b} & 64.3 & 69.3 & $> 16.6$ \\
\hline 
061110 & 43.5 & 48.5 &  $>16.4$ \\
\hline 
061222 & 47.2 & 52.2 &  $> 17.0$ \\
061222 & 54.2 & 59.2 &  $> 16.9$ \\
061222 & 61.2 & 66.2 &  $> 17.0$ \\
061222 & 68.2 & 73.2 &  $> 16.9$ \\
061222 & 75.2 & 80.2 &  $> 16.9$ \\
061222 & 82.2 & 87.2 &  $> 16.9$ \\
061222 & 89.2 & 94.2 &  $> 17.0$ \\
061222 & 96.2 & 101.2 &  $> 16.9$ \\
061222 & 103.2 & 108.2 &  $> 16.9$ \\
061222 & 110.1 & 115.1 &  $> 17.0$ \\
\enddata
\tablecomments{All times are in seconds since the burst onset, which are (UT): 03:33:12 UT (GRB\,050306), 04:29:02.4 (GRB\,050713A),  03:49:29 (GRB\,050822), 11:22:42 (GRB\,050915A), 15:02:00 (GRB\,050922B), 11:11:36.2 (GRB\,051001), 01:36:12.8 (GRB\,060312), 02:27:52 (GRB\,060515), 12:43:48.5 (GRB\,060614), 19:12:29.2 (GRB\,060729), 11:47:21.3 (GRB\,061110), 03:28:52.1 (GRB\,061222). Magnitudes are quoted without correction for local or Galactic extinction, and are $R$-equivalent unfiltered values. The extinction corrections are (in $A_R$ magnitudes): 1.817
(GRB\,050306), 1.107 (GRB\,050713A), 0.04 (GRB\,050822), 0.07
(GRB\,050915A), 0.098 (GRB\,050922B), 0.04 (GRB\,051001), 0.472
(GRB\,060312), 0.073 (GRB\,060515), 0.058 (GRB\,060614), 0.145
(GRB\,060729), 0.242 (GRB\,061110) and 0.266 (GRB\,061222).}
\tablenotetext{a}{The {\it Swift} UVOT detected the OT in this event during an exposure from 102--202 sec post-trigger \citep{GCN5252}. The ROTSE limits are consistent with the more sensitive UVOT detection.}
\tablenotetext{b}{GRB\,060729 was promptly detected, however, the {\em first} 5 sec observation only yielded a limit for the OT.}
\end{deluxetable}

\begin{deluxetable}{lcccccccc}
\tablewidth{0pt}
\tablecaption{Spectral Indices $\beta_{\mathrm{OPT}-\gamma}$ (or $\beta_{\mathrm{OPT}-X}$) from Prompt Optical Detections\label{tab:optgam1}}
\tabletypesize{\scriptsize}
\tablehead{
  \colhead{GRB} &
  \colhead{$t_{\mathrm{start}}$} &
  \colhead{$t_{\mathrm{end}}$} &
  \colhead{Band} &
  \colhead{$f_{\nu}\mathrm{(OPT)}$} &
  \colhead{$\nu_{\gamma}$ [or $\nu_{X}$]} &
  \colhead{$f_{\nu}$($\gamma$) [or $f_{\nu}$($X$)]} &
  \colhead{$\beta_{\gamma}$ [or $\beta_{X}$]} &
  \colhead{$\beta_{\mathrm{OPT}-\gamma}$}\\
  \colhead{ } &
  \colhead{(s)} &
  \colhead{(s)} &
  \colhead{} &
  \colhead{(mJy)} &
  \colhead{($10^{18}$Hz)} &
  \colhead{($\mu$Jy)} &
  \colhead{} &
  \colhead{[or $\beta_{\mathrm{OPT}-X}$]}
}
\startdata
050820A & 252 & 282 & $C_{R}$ & 2.612 $\pm$ 0.058 & 25 & 453 $\pm$ 17 & -0.371 $\pm$ 0.061 & -0.161 $\pm$ 0.004 \\
050820A & 402 & 432 & $C_{R}$ & 4.814 $\pm$ 0.084 & 25 & 314 $\pm$ 16 & -0.415 $\pm$ 0.078 & -0.251 $\pm$ 0.005 \\
050820A & 515 & 545 & $C_{R}$ & 4.452 $\pm$ 0.077 & 27 & 138 $\pm$ 15 & -0.707 $\pm$ 0.143 & -0.32 $\pm$ 0.01 \\
060111B & 58.0 & 63.0 & $C_{R}$ & 5.97 $\pm$ 0.66 & 14 & 89 $\pm$ 14 & -1.02 $\pm$ 0.20 & -0.41 $\pm$ 0.02 \\
060218 & 691 & 1027 & $C_{R}$ & 0.254 $\pm$ 0.026 & 11 & 91.3 $\pm$ 7.4 & -1.5 $\pm$ 0.1 & -0.10 $\pm$ 0.01 \\
060218 & 700 & 1000 & $V$ & 0.106 $\pm$ 0.020 & 11 & 91.3 $\pm$ 7.4 & -1.5 $\pm$ 0.1 & -0.02 $\pm$ 0.02 \\
060729 & 73.4 & 83.4 & $C_{R}$ & 0.68 $\pm$ 0.19 & 16 & 203 $\pm$ 20 & -0.611 $\pm$ 0.093 & -0.11 $\pm$ 0.04 \\
060729 & 92.9 & 97.9 & $C_{R}$ & 1.90 $\pm$ 0.18 & 14 & 173 $\pm$ 15 & -0.986 $\pm$ 0.087 & -0.23 $\pm$ 0.02 \\
060729 & 114.8 & 119.8 & $C_{R}$ & 0.74 $\pm$ 0.18 & 15 & 35.4 $\pm$ 6.9 & -0.896 $\pm$ 0.065 & -0.29 $\pm$ 0.04 \\
060729 & 128.8 & 133.8 & $C_{R}$ & 0.61 $\pm$ 0.18 & 15 & 20.0 $\pm$ 6.3 & -0.896 $\pm$ 0.065 & -0.33 $\pm$ 0.04 \\
060729 & .. & .. & .. & ..  & 0.67 & 1527 $\pm$ 34 & -2.004 $\pm$ 0.029 & 0.13 $\pm$ 0.04 \\
060904B & 18.5 & 31.5 & $C_{R}$ & 0.278 $\pm$ 0.055 & 18 & 20.52 $\pm$ 6.3 & -0.416 $\pm$ 0.081 & -0.288 $\pm$ 0.035 \\ 
060904B & 146.4 & 166.4 & $C_{R}$ & 0.370 $\pm$ 0.072 & 13 & 36.2 $\pm$ 7.5 & -1.30 $\pm$ 0.18 & -0.270 $\pm$ 0.046 \\
060904B & .. & .. & .. & .. & 0.67 & 1011 $\pm$ 29 & -1.26 $\pm$ 0.03 & 0.080 $\pm$ 0.027 \\
060927 & 16.8 & 21.8 & $C_{i}$\tablenotemark{a} & 6.1 $\pm$ 1.1 & 16 & 125 $\pm$ 16 & -0.77 $\pm$ 0.13 & -0.365 $\pm$ 0.030 \\
061007 & 27.2 & 32.2. & $C_{R}$ & 10.83 $\pm$ 0.69 & 21 & 3198 $\pm$ 54 & 0.163 $\pm$ 0.028 & -0.114 $\pm$ 0.008 \\
061007 & 41.0 & 46.0 & $C_{R}$ & 286.9 $\pm$ 4.9 & 20 & 1849 $\pm$ 29 & 0.103 $\pm$ 0.026 & -0.472 $\pm$ 0.003 \\
061007 & 55.4 & 60.4 & $C_{R}$ & 481.0 $\pm$ 9.2 & 20 & 2776 $\pm$ 31 & 0.054 $\pm$ 0.020 & -0.483 $\pm$ 0.003 \\
061007 & 77.8 & 82.8 & $C_{R}$ & 407.4 $\pm$ 5.7 & 16 & 215.9 $\pm$ 12 & -0.673 $\pm$ 0.072 & -0.72 $\pm$ 0.01 \\
061007 & 92.0 & 97.0 & $C_{R}$ & 500.5 $\pm$ 7.8 & 0.67 & 1400.4 $\pm$ 8.2 & -0.906 $\pm$ 0.013 & -0.810 $\pm$ 0.002 \\
061007 & .. & .. & .. & .. & 15 & 92  $\pm$ 12 & -0.824 $\pm$ 0.093 & -0.828 $\pm$  0.013 \\
061007 & 106 & 111 & $C_{R}$ & 449.0 $\pm$ 6.4 & 0.67 & 1118.0 $\pm$ 6.0 & -0.906 $\pm$ 0.013 & -0.826 $\pm$ 0.002 \\
061007 & .. & .. & .. & .. & 15 & 33.4 $\pm$ 7.6 & -0.824 $\pm$ 0.093 & -0.916 $\pm$  0.025 \\
061007 & 120 & 125 & $C_{R}$ & 376.6 $\pm$ 9.5 & 0.67 & 909.9 $\pm$ 4.5 & -0.906 $\pm$ 0.013 & -0.830 $\pm$ 0.004 \\
061007 & .. & .. & .. & .. & 15 & 39.6 $\pm$ 7.8 & -0.824 $\pm$ 0.093 & -0.882 $\pm$  0.023 \\
061007 & 135 & 140 & $C_{R}$ & 333.7 $\pm$ 5.0 & 0.67 & 760.5 $\pm$ 3.5 & -0.906 $\pm$ 0.013 & -0.838 $\pm$ 0.002 \\
061007 & .. & .. & .. & .. & 15 & 26.0 $\pm$ 7.2 & -0.824 $\pm$ 0.093 & -0.911 $\pm$  0.030 \\
061007 & 149 & 154 & $C_{R}$ & 280.9 $\pm$ 4.9 & 0.67 & 644.8 $\pm$ 2.8 & -0.906 $\pm$ 0.013 & -0.837 $\pm$ 0.002 \\
061007 & .. & .. & .. & .. & 15 & 24.9 $\pm$ 7.1 & -0.824 $\pm$ 0.093 & -0.899 $\pm$  0.031 \\
061007 & 164 & 169 & $C_{R}$ & 233.5 $\pm$ 4.7 & 0.67 & 554.0 $\pm$ 2.3 & -0.906 $\pm$ 0.013 & -0.832 $\pm$ 0.003 \\
061007 & .. & .. & .. & .. & 15 & 21.6 $\pm$ 2.6 & -0.824 $\pm$ 0.093 & -0.895 $\pm$  0.014 \\
061007 & 178 & 198 & $C_{R}$ & 183.4 $\pm$ 3.6 & 0.67 & 452.6 $\pm$ 1.8 & -0.906 $\pm$ 0.013 & -0.827 $\pm$ 0.003 \\
061007 & .. & .. & .. & .. & 15 & 13.6 $\pm$ 3.6 & -0.824 $\pm$ 0.093 & -0.916 $\pm$  0.029 \\
061007 & 207 & 227 & $C_{R}$ & 149.4 $\pm$ 2.4 & 0.67 & 357.0 $\pm$ 1.4 & -0.906 $\pm$ 0.013 & -0.831 $\pm$ 0.002 \\
061007 & .. & .. & .. & .. & 15 & 13.0 $\pm$ 3.5 & -0.824 $\pm$ 0.093 & -0.901 $\pm$  0.030 \\
061007 & 237 & 257 & $C_{R}$ & 127.7 $\pm$ 1.6 & 0.67 & 289.9 $\pm$ 1.1 & -0.906 $\pm$ 0.013 & -0.839 $\pm$ 0.002 \\
061007 & .. & .. & .. & .. & 15 & 10.4 $\pm$ 1.8 & -0.824 $\pm$ 0.093 & -0.906 $\pm$  0.019 \\
061121 & 21.7 & 69.5 & $C_{R}$ & 0.86 $\pm$ 0.54 & 18 & 263.2 $\pm$ 4.3 & -0.403 $\pm$ 0.027 & -0.11 $\pm$ 0.06 \\
061121 & 78.3 & 83.3 & $C_{R}$ & 3.33 $\pm$ 0.94 & 16 & 328 $\pm$ 13  & -0.668 $\pm$ 0.053 & -0.22 $\pm$ 0.03 \\
061121 & .. & .. & .. & .. & 0.24 & 6930 $\pm$ 320  & -0.07 $\pm$ 0.08\tablenotemark{b} & 0.118 $\pm$ 0.046 \\
061121 & 92.5 & 126 & $C_{R}$ & 1.04 $\pm$ 0.51 & 15 & 49.3 $\pm$ 5.0  & -0.83 $\pm$ 0.10 & -0.29 $\pm$ 0.05 \\
061121 & .. & .. & .. & .. & 0.24 & 2024 $\pm$ 74  & -\tablenotemark{c} & 0.079 $\pm$ 0.091 \\
061126 & 20.9 & 25.9 & $C_{R}$ & 60.65 $\pm$ 0.55 & 19  & 473 $\pm$ 16  & -0.262 $\pm$ 0.059 & -0.459 $\pm$ 0.007 \\
061126 & 29.8 & 34.8 & $C_{R}$ & 41.96 $\pm$ 0.77 & 19  & 65.9 $\pm$ 9.2  & .. & -0.61 $\pm$ 0.02 \\
061126 & 38.6 & 43.6 & $C_{R}$ & 28.50 $\pm$ 0.78 & 19  & 47 $\pm$ 12  & .. & -0.60 $\pm$ 0.03 \\
\enddata
\tablecomments{Optical and $\gamma$-ray flux densities and spectral indices correspond to the time intervals $t_{\mathrm{start}}$ -- $t_{\mathrm{end}}$ from the GRB trigger. The sources of the data are discussed in \S\ref{sec:detdata}, and the optical data are corrected for Galactic extinction.}
\tablenotetext{a}{Filterless observations of this high-{\it z} event were calibrated to the flux density at 819 nm, approximately {\it i}-band, see Ruiz-Velasco
{\it et al.}, in prep.}
\tablenotetext{b}{Taken from the spread in spectral indices with different extinction models, see \citet{page07}, Table 4.}
\tablenotetext{c}{There is no value given by \citet{page07} for the spectral shape during the steep decline from the peak. For many cases, the steep X-ray phase has been reported as spectrally indistinguishable from the later shallow decay, but in some cases $\beta_{X}$ is softer during the initial rapid decay \citep{nkgpg06}.}
\end{deluxetable}

\begin{deluxetable}{lcccccccccc}
\tablewidth{0pt}
\tablecaption{Spectral Index $\beta_{\mathrm{OPT}-\gamma}$ Limits from Prompt Optical Limits\label{tab:optgam2}}
\tabletypesize{\scriptsize}
\tablehead{
  \colhead{GRB} &
  \colhead{$t_{\mathrm{start}}$} &
  \colhead{$t_{\mathrm{end}}$} &
  \colhead{Band} &
  \colhead{$f_{\nu}$(OPT)} &
  \colhead{$\nu_{\gamma}$} &
  \colhead{$f_{\nu}$($\gamma$)} &
  \colhead{$\beta_{\gamma}$} &
  \colhead{$\beta_{\mathrm{OPT}-\gamma}$} \\
  \colhead{} &
  \colhead{(s)} &
  \colhead{(s)} &
  \colhead{} &
  \colhead{(mJy, Limit)} &
  \colhead{10$^{18}$Hz} &
  \colhead{$\mu$Jy} &
  \colhead{} &
  \colhead{(Limit)} 
}
\startdata
050306 & 64.8 & 69.8 & $C_{R}$ & $<$\,10.6 & 17 & 136 $\pm$ 24 & -0.477 $\pm$ 0.042 & $>$\,-0.435 \\
050306 & 78.9 & 83.9 & $C_{R}$ & $<$\,7.6 & 17 & 416 $\pm$ 27 & -0.477 $\pm$ 0.042 & $>$\,-0.284 \\
050306 & 93.5 & 98.5 & $C_{R}$ & $<$\,7.8 & 17 & 259 $\pm$ 25 & -0.477 $\pm$ 0.042 & $>$\,-0.336 \\
050306 & 108.3 & 113.3 & $C_{R}$ & $<$\,8.7 & 17 & 347 $\pm$ 27 & -0.477 $\pm$ 0.042 & $>$\,-0.316 \\
050306 & 122.7 & 127.7 & $C_{R}$ & $<$\,8.7 & 17 & 115 $\pm$ 23 & -0.477 $\pm$ 0.042 & $>$\,-0.434 \\
050306 & 180.4 & 185.4 & $C_{R}$ & $<$\,8.0 & 17 & 176 $\pm$ 25 & -0.477 $\pm$ 0.042 & $>$\,-0.379 \\
050713A & 72.1 & 77.1 & $C_{R}$ & $<$\,2.1 & 15 & 19.2$^{+4.8}_{-8.7}$ & -0.85 $\pm$ 0.16 & $>$\,-0.51 \\
050713A & 104.7 & 124.7 & $C_{R}$ & $<$\,1.2 & 15 & 18.5$^{+2.2}_{-6.4}$ & -0.85 $\pm$ 0.16 & $>$\,-0.44 \\
050822 & 31.8 & 36.8 & $C_{R}$ & $<$\,1.9 & 12 & 42 $\pm$ 16 & -1.32 $\pm$ 0.09 & $>$\,-0.43 \\
050822 & 39.8 & 44.8 & $C_{R}$ & $<$\,1.9 & 11 & 178$^{+19}_{-67}$ & -1.61 $\pm$ 0.20 & $>$\,-0.28 \\
050822 & 47.8 & 52.8 & $C_{R}$ & $<$\,2.0 & 13 & 194$^{+12}_{-45}$ & -1.28 $\pm$ 0.13 & $>$\,-0.25 \\
050822 & 55.9 & 60.9 & $C_{R}$ & $<$\,1.9 & 12 & 181$^{+12}_{-41}$ & -1.41 $\pm$ 0.13 & $>$\,-0.26 \\
050822 & 63.9 & 68.9 & $C_{R}$ & $<$\,2.1 & 12 & 38.4 $\pm$ 9.9 & -1.32 $\pm$ 0.09 & $>$\,-0.43 \\
050822 & 95.9 & 100.9 & $C_{R}$ & $<$\,1.9 & 12 & 23.2 $\pm$ 8.9 & -1.32 $\pm$ 0.09 & $>$\,-0.48 \\
050915A & 42.9 & 47.9 & $C_{R}$ & $<$\,0.54 & 18 & 32.2$^{+6.2}_{-10.7}$ & -0.38 $\pm$ 0.10 & $>$\,-0.31 \\
050922B & 258.4 & 263.4 & $C_{R}$ & $<$\,0.94 & 14 & 39.5$^{+6.7}_{-12.0}$ & -0.99 $\pm$ 0.12 & $>$\,-0.34 \\
050922B & 273.0 & 278.0 & $C_{R}$ & $<$\,0.87 & 14 & 19.7 $\pm$ 7.5 & -0.99 $\pm$ 0.12 & $>$\,-0.42 \\
051001 & 85.7 & 90.7 & $C_{R}$ & $<$\,1.0 & 14 & 17.4$^{+2.5}_{-4.1}$ & -1.06 $\pm$ 0.10 & $>$\,-0.42 \\
051001 & 100.1 & 105.1 & $C_{R}$ & $<$\,0.98 & 14 & 27.5$^{+3.2}_{-5.8}$ & -1.06 $\pm$ 0.10 & $>$\,-0.37 \\
051001 & 114.3 & 119.3 & $C_{R}$ & $<$\,1.0 & 14 & 28.5 $\pm$ 7.2 & -1.06 $\pm$ 0.10 & $>$\,-0.38 \\
051001 & 128.6 & 133.6 & $C_{R}$ & $<$\,0.98 & 14 & 58.2$^{+6.1}_{-11.6}$ & -1.06 $\pm$ 0.10 & $>$\,-0.30 \\
051001 & 143.1 & 148.1 & $C_{R}$ & $<$\,0.99 & 14 & 57.0$^{+6.0}_{-11.5}$ & -1.06 $\pm$ 0.10 & $>$\,-0.30 \\
051001 & 157.6 & 162.6 & $C_{R}$ & $<$\,1.0 & 14 & 48.4 $\pm$ 8.4 & -1.06 $\pm$ 0.10 & $>$\,-0.32 \\
051001 & 172.3 & 177.3 & $C_{R}$ & $<$\,1.1 & 14 & 17.8 $\pm$ 3.6 & -1.06 $\pm$ 0.10 & $>$\,-0.43 \\
051001 & 186.9 & 191.9 & $C_{R}$ & $<$\,1.1 & 14 & 18.5 $\pm$ 6.7 & -1.06 $\pm$ 0.10 & $>$\,-0.45 \\
060312 & 20.3 & 25.3 & $C_{R}$ & $<$\,11.2 & 16 & 27.2 $\pm$ 7.1 & -0.772 $\pm$ 0.054 & $>$\,-0.610 \\
060312 & 27.4 & 32.4 & $C_{R}$ & $<$\,11.0 & 16 & 20.7 $\pm$ 3.4 & -0.772 $\pm$ 0.054 & $>$\,-0.622 \\
060312 & 34.4 & 39.4 & $C_{R}$ & $<$\,10.0 & 16 & 14.3 $\pm$ 1.3 & -0.772 $\pm$ 0.054 & $>$\,-0.653 \\
060312 & 41.5 & 46.5 & $C_{R}$ & $<$\,9.2 & 16 & 21.7 $\pm$ 7.0 & -0.772 $\pm$ 0.054 & $>$\,-0.621 \\
060312 & 48.7 & 53.7 & $C_{R}$ & $<$\,9.1 & 16 & 12.0 $\pm$ 3.4 & -0.772 $\pm$ 0.054 & $>$\,-0.671 \\
060515 & 58.8 & 63.8 & $C_{R}$ & $<$\,5.1 & 19 & 30.5$^{+3.6}_{-8.8}$ & -0.26 $\pm$ 0.14 & $>$\,-0.52 \\
060614 & 26.8 & 31.8 & $C_{R}$ & $<$\,1.7 & 14 & 849 $\pm$ 25 & -1.103 $\pm$ 0.026 & $>$\,-0.068 \\
060614 & 40.6 & 45.6 & $C_{R}$ & $<$\,1.8 & 14 & 913 $\pm$ 22 & -1.103 $\pm$ 0.026 & $>$\,-0.071 \\
060614 & 55.2 & 60.2 & $C_{R}$ & $<$\,1.8 & 14 & 423 $\pm$ 12 & -1.103 $\pm$ 0.026 & $>$\,-0.144 \\
060614 & 69.6 & 74.6 & $C_{R}$ & $<$\,1.9 & 13 & 300 $\pm$ 13 & -1.254 $\pm$ 0.045 & $>$\,-0.184 \\
060614 & 83.9 & 88.9 & $C_{R}$ & $<$\,1.8 & 13 & 256 $\pm$ 12 & -1.254 $\pm$ 0.045 & $>$\,-0.197 \\
060614 & 98.3 & 103.3 & $C_{R}$ & $<$\,1.8 & 13 & 197 $\pm$ 10 & -1.254 $\pm$ 0.045 & $>$\,-0.225 \\
060614 & 112.6 & 117.6 & $C_{R}$ & $<$\,1.8 & 13 & 73.6 $\pm$ 6.9 & -1.254 $\pm$ 0.045 & $>$\,-0.326 \\
060614 & 126.8 & 131.8 & $C_{R}$ & $<$\,1.8 & 13 & 37.8 $\pm$ 6.2 & -1.254 $\pm$ 0.045 & $>$\,-0.399 \\
060614 & 140.7 & 145.7 & $C_{R}$ & $<$\,1.8 & 13 & 51.0 $\pm$ 6.4 & -1.254 $\pm$ 0.045 & $>$\,-0.364 \\
060614 & 155.2 & 160.2 & $C_{R}$ & $<$\,1.9 & 13 & 34.4 $\pm$ 6.1 & -1.254 $\pm$ 0.045 & $>$\,-0.411 \\
060614 & 169.2 & 189.2 & $C_{R}$ & $<$\,1.0 & 13 & 12.3 $\pm$ 3.0 & -1.254 $\pm$ 0.045 & $>$\,-0.463 \\
060729 & 64.3 & 69.3 & $C_{R}$ & $<$\,0.81 & 17 & 43 $\pm$ 15 & -0.517 $\pm$ 0.095 & $>$\,-0.254 \\
061110 &  43.5 & 48.5 & $C_{R}$ & $<$\,1.11 & 16 & 22.9 $\pm$ 7.6 & -0.654 $\pm$ 0.087 & $>$\,-0.416 \\
061222 &  47.2 &  52.2 & $C_{R}$ & $<$\,0.65 & 17 & 38.5 $\pm$ 9.4 & -0.487 $\pm$ 0.095 & $>$\,-0.303 \\
061222 &  54.2 &  59.2 & $C_{R}$ & $<$\,0.67 & 18 & 181.4$^{+5.3}_{-28.0}$ & -0.365 $\pm$ 0.093 & $>$\,-0.139 \\
061222 &  61.2 &  66.2 & $C_{R}$ & $<$\,0.63 & 16 & 147.221$^{+5.0}_{-23.8}$ & -0.76 $\pm$ 0.10 & $>$\,-0.157 \\
061222 &  68.2 &  73.2 & $C_{R}$ & $<$\,0.65 & 18 & 196.9$^{+5.4}_{-24.1}$ & -0.280 $\pm$ 0.079 & $>$\,-0.126 \\
061222 &  75.2 &  80.2 & $C_{R}$ & $<$\,0.70 & 17 & 227.7$^{+4.6}_{-18.6}$ & -0.537 $\pm$ 0.065 & $>$\,-0.115 \\
061222 &  82.2 &  87.2 & $C_{R}$ & $<$\,0.70 & 20 & 1309 $\pm$ 25 & -0.002 $\pm$ 0.025 & $>$\,0.058 \\
061222 &  89.2 &  94.2 & $C_{R}$ & $<$\,0.63 & 17 & 580.8$^{+8.6}_{-19.7}$ & -0.446 $\pm$ 0.035  & $>$\,-0.011 \\
061222 &  96.2 & 101.2 & $C_{R}$ & $<$\,0.66 & 15 & 91.1$^{+5.3}_{-17.7}$ & -0.84 $\pm$ 0.11 & $>$\,-0.212 \\
061222 & 103.2 & 108.2 & $C_{R}$ & $<$\,0.68 & 15 & 47.8$^{+5.5}_{-11.8}$ & -0.84 $\pm$ 0.11 & $>$\,-0.284 \\
061222 & 110.1 & 115.1 & $C_{R}$ & $<$\,0.65 & 15 & 25.1 $\pm$ 7.1 & -0.84 $\pm$ 0.11 & $>$\,-0.354 \\
\enddata
\tablecomments{Optical flux limits and $\gamma$-ray flux densities $f_{\nu}$ and their spectral indices, corresponding to the time intervals $t_{\mathrm{start}}$ -- $t_{\mathrm{end}}$ from the GRB trigger. $\gamma$-ray count rates were all detected at the 3\,$\sigma$ level or better, although the spectral fits for some cases result in $f_{\nu}$ with signal-to-noise formally $< 3$.  The optical limits are from Table \ref{tab:optlim}, corrected for Galactic extinction.}
\end{deluxetable}

\begin{thebibliography}{}

\bibitem[\protect\astroncite{{Akerlof} {\rm et~al.\/}}{2003}]{akmrs03}
 {Akerlof}, C.~W., et~al.
\newblock Jan. 2003, \pasp, 115, 132

\bibitem[\protect\astroncite{{Barthelmy} {\rm et~al.\/}}{2006}]{GCN5256}
 {Barthelmy}, S., et~al.
\newblock 2006, GCN Circ. No. 5256

\bibitem[\protect\astroncite{{Bertin} \& {Arnouts}}{1996}]{ba96}
{Bertin}, E. \& {Arnouts}, S.
\newblock June 1996, \aaps, 117, 393

\bibitem[\protect\astroncite{{Bessell}}{1979}]{b79}
{Bessell}, M.~S.
\newblock Oct. 1979, \pasp, 91, 589

\bibitem[\protect\astroncite{{Butler} {\rm et~al.\/}}{2007}]{bkbc07}
{Butler}, N.~R., {Kocevski}, D., {Bloom}, J.~S., \& {Curtis}, J.~L.
\newblock June 2007, astro-ph/0706.1275

\bibitem[\protect\astroncite{{Campana} {\rm et~al.\/}}{2006}]{cmbbb06}
 {Campana}, S., et~al.
\newblock Aug. 2006, \nat, 442, 1008

\bibitem[\protect\astroncite{{Chen} {\rm et~al.\/}}{2006}]{GCN5797}
{Chen}, Y.~C., {Lee}, Y.~H., {Huang}, K.~Y., {Ip}, W.~H., \& {Urata}, Y.
\newblock 2006, GCN Circ. No. 5797

\bibitem[\protect\astroncite{{D'Avanzo} {\rm et~al.\/}}{2005}]{GCN3089}
{D'Avanzo}, P., {Fugazza}, D., {Covino}, S., {Malesani}, D., {Masetti}, N.,
  {Palazzi}, E., {Antonelli}, L.~A., {Israel}, G.~L., \& {Andreuzzi}, G.
\newblock 2005, GCN Circ. No. 3089

\bibitem[\protect\astroncite{{Fishman} \& {Meegan}}{1995}]{fm95}
{Fishman}, G.~J. \& {Meegan}, C.~A.
\newblock 1995, \araa, 33, 415

\bibitem[\protect\astroncite{{Fynbo} {\rm et~al.\/}}{2006}]{GCN5651}
{Fynbo}, J.~P.~U., {Jakobsson}, P., {Jensen}, B.~L., {Hjorth}, J., {Sollerman},
  J., {Watson}, D., {Cerón}, J.~M.~C., {Vreeswijk}, P., \& {Andersen}, M.~I.
\newblock 2006, GCN Circ. No. 5651

\bibitem[\protect\astroncite{{Gehrels} {\rm et~al.\/}}{2004}]{gcgmn04}
 {Gehrels}, N., et~al.
\newblock Aug. 2004, \apj, 611, 1005

\bibitem[\protect\astroncite{{Kumar} {\rm et~al.\/}}{2007}]{kmpwo07}
 {Kumar}, P., et~al.
\newblock Mar. 2007, \mnras, 376, L57

\bibitem[\protect\astroncite{{Malesani} {\rm et~al.\/}}{2005}]{GCN3582}
{Malesani}, D., {D'Avanzo}, P., {Palazzi}, E., {Israel}, G.~L., {Chincarini},
  G., {Stella}, L., \& {Pedani}, M.
\newblock 2005, GCN Circ. No. 3582

\bibitem[\protect\astroncite{{Mangano} {\rm et~al.\/}}{2006}]{GCN5254}
{Mangano}, V., {Parola}, V.~L., {Troja}, E., {Cusumano}, G., {Mineo}, T.,
  {Parsons}, A., \& {Kennea}, J.
\newblock 2006, GCN Circ. No. 5254

\bibitem[\protect\astroncite{{Medvedev}}{2006}]{medved06}
{Medvedev}, M.~V.
\newblock Feb. 2006, \apj, 637, 869

\bibitem[\protect\astroncite{{Meszaros}}{2006}]{m06}
{Meszaros}, P.
\newblock 2006, Reports of Progress in Physics, 69, 2259

\bibitem[\protect\astroncite{{Meszaros} {\rm et~al.\/}}{1994}]{mrp94}
{Meszaros}, P., {Rees}, M.~J., \& {Papathanassiou}, H.
\newblock Sept. 1994, \apj, 432, 181

\bibitem[\protect\astroncite{{Mundell} {\rm et~al.\/}}{2007}]{mmgks07}
 {Mundell}, C.~G., et~al.
\newblock May 2007, \apj, 660, 489

\bibitem[\protect\astroncite{{Nousek} {\rm et~al.\/}}{2006}]{nkgpg06}
 {Nousek}, J.~A., et~al.
\newblock May 2006, \apj, 642, 389

\bibitem[\protect\astroncite{{Nysewander} \& {Haislip}}{2006}]{GCN5236}
{Nysewander}, M. \& {Haislip}, J.
\newblock 2006, GCN Circ. No. 5236

\bibitem[\protect\astroncite{{O'Brien} {\rm et~al.\/}}{2006}]{owogpv06}
 {O'Brien}, P.~T., et~al.
\newblock Aug. 2006, \apj, 647, 1213

\bibitem[\protect\astroncite{{Page} {\rm et~al.\/}}{2007}]{page07}
 {Page}, K.~L., et~al.
\newblock 2007, submitted to ApJ

\bibitem[\protect\astroncite{{Parsons} {\rm et~al.\/}}{2006}]{GCN5252}
 {Parsons}, A.~M., et~al.
\newblock 2006, GCN Circ. No. 5252

\bibitem[\protect\astroncite{{Pe'er} \& {Waxman}}{2004}]{pw04}
{Pe'er}, A. \& {Waxman}, E.
\newblock Sept. 2004, \apj, 613, 448

\bibitem[\protect\astroncite{{Perley} {\rm et~al.\/}}{2007}]{pbbph07}
 {Perley}, D.~A., et~al.
\newblock Mar. 2007, ApJ submitted, astro-ph/0703538

\bibitem[\protect\astroncite{{Piran}}{2005}]{p05}
{Piran}, T.
\newblock 2005, Reviews of Modern Physics, 76, 1143

\bibitem[\protect\astroncite{{Price} {\rm et~al.\/}}{2006}]{GCN5275}
{Price}, P.~A., {Berger}, E., \& {Fox}, D.~B.
\newblock 2006, GCN Circ. No. 5275

\bibitem[\protect\astroncite{{Quimby} \& {Rykoff}}{2006}]{GCN5377}
{Quimby}, R. \& {Rykoff}, E.~S.
\newblock 2006, GCN Circ. No. 5377

\bibitem[\protect\astroncite{{Quimby} {\rm et~al.\/}}{2006a}]{qryaa05}
 {Quimby}, R.~M., et~al.
\newblock 2006a, \apj, 640, 402

\bibitem[\protect\astroncite{{Quimby} {\rm et~al.\/}}{2006b}]{GCN5366}
{Quimby}, R., {Swan}, H., {Rujopakarn}, W., \& {Smith}, D.~A.
\newblock 2006b, GCN Circ. No. 5366

\bibitem[\protect\astroncite{{Quimby} {\rm et~al.\/}}{2006c}]{GCN4782}
{Quimby}, R., {Schaefer}, B.~E., \& {Swan}, H.
\newblock 2006c, GCN Circ. No. 4782

\bibitem[\protect\astroncite{{Rykoff} \& {Rujopakarn}}{2006}]{GCN5706}
{Rykoff}, E.~S. \& {Rujopakarn}, W.
\newblock 2006, GCN Circ. No. 5706

\bibitem[\protect\astroncite{{Rykoff} {\rm et~al.\/}}{2006}]{GCN5504}
{Rykoff}, E.~S., {Rujopakarn}, W., \& {Yuan}, F.
\newblock 2006, GCN Circ. No. 5504

\bibitem[\protect\astroncite{{Rykoff} {\rm et~al.\/}}{2004}]{rspaa04}
 {Rykoff}, E.~S., et~al.
\newblock Feb. 2004, \apj, 601, 1013

\bibitem[\protect\astroncite{{Rykoff} {\rm et~al.\/}}{2005}]{rykaa05}
 {Rykoff}, E.~S., et~al.
\newblock Oct. 2005, \apjl, 631, L121

\bibitem[\protect\astroncite{{Schaefer} {\rm et~al.\/}}{2006}]{srsq06}
{Schaefer}, B.~E., {Rykoff}, E.~S., {Smith}, D.~A., \& {Quimby}, R.
\newblock 2006, GCN Circ. No. 5222

\bibitem[\protect\astroncite{{Schaefer} \& {Xiao}}{2006}]{sx06}
{Schaefer}, B.~E. \& {Xiao}, L.
\newblock Aug. 2006, ApJL submitted, astro-ph/0608441

\bibitem[\protect\astroncite{{Schlegel} {\rm et~al.\/}}{1998}]{sfd98}
{Schlegel}, D.~J., {Finkbeiner}, D.~P., \& {Davis}, M.
\newblock June 1998, \apj, 500, 525

\bibitem[\protect\astroncite{{Stetson}}{1987}]{stetson87}
{Stetson}, P.~B.
\newblock Mar. 1987, \pasp, 99, 191

\bibitem[\protect\astroncite{{Thoene} {\rm et~al.\/}}{2006}]{GCN5373}
{Thoene}, C.~C., {Levan}, A., {Jakobsson}, P., {Rol}, E., {Gorosabel}, J.,
  {Jensen}, B.~L., {Hjorth}, J., \& {Vreeswijk}, P.
\newblock 2006, GCN Circ. No. 5373

\bibitem[\protect\astroncite{{Thompson}}{1994}]{thompson94}
{Thompson}, C.
\newblock Oct. 1994, \mnras, 270, 480

\bibitem[\protect\astroncite{{Usov}}{1994}]{usov94}
{Usov}, V.~V.
\newblock Apr, 1994, \mnras, 267, 1035

\bibitem[\protect\astroncite{{Vestrand} {\rm et~al.\/}}{2005}]{vwwfs05}
 {Vestrand}, W.~T., et~al.
\newblock May 2005, \nat, 435, 178

\bibitem[\protect\astroncite{{Vestrand} {\rm et~al.\/}}{2006}]{vwwag06}
 {Vestrand}, W.~T., et~al.
\newblock July 2006, \nat, 442, 172

\bibitem[\protect\astroncite{{Wei}}{2007}]{wei07}
{Wei}, D.~M.
\newblock Jan. 2007, \mnras, 374, 525

\bibitem[\protect\astroncite{{Yost} {\rm et~al.\/}}{2007}]{ysraa07}
 {Yost}, S.~A., et~al.
\newblock 2007, \apj, 657, 925

\bibitem[\protect\astroncite{{Yost} {\rm et~al.\/}}{2006}]{GCN4488}
{Yost}, S.~A., {Yuan}, F., {Swan}, H., \& {Akerlof}, C.
\newblock 2006, GCN Circ. No. 4488

\bibitem[\protect\astroncite{{Zhang}}{2007}]{Zhang07}
{Zhang}, B.
\newblock Feb. 2007, ChJAA, 7, 1

\end{thebibliography}
\end{document}